\documentclass[11pt]{article}
\usepackage{graphicx} 

\usepackage{fancyhdr}
\usepackage{isomath}
\usepackage{mathtools} 
\usepackage{amsbsy}
\usepackage{amssymb}
\usepackage{amscd,amsfonts}
\usepackage{bigints}
\usepackage{graphicx}
\usepackage{verbatim}
\usepackage{euscript}
\usepackage{alltt}
\usepackage{stmaryrd}
\usepackage{relsize}
\usepackage{enumerate}
\usepackage{url}
\usepackage[stable]{footmisc}
\usepackage{breakurl}
\usepackage{hyperref}
\usepackage{comment}

\usepackage[maxbibnames=9,sorting=none]{biblatex}
\usepackage[font=small,labelfont=bf]{caption}
\usepackage[font=small,labelfont=bf]{subcaption}
\usepackage{float}
\usepackage{mwe}
\usepackage{algorithm}
\usepackage{algpseudocode}
\usepackage{xcolor}
\usepackage[super]{nth}
\usepackage{soul}

\DeclareGraphicsExtensions{.eps,.pdf}
\usepackage{amsmath}
\usepackage{amsbsy}
\usepackage{amssymb}
\usepackage{amscd}
\usepackage{amsfonts}
\usepackage{isomath}

\newcommand{\R}{\mathbb R}

\newcommand{\beq}{\begin{equation}}
\newcommand{\eeq}{\end{equation}}
\newcommand{\beqs}{\begin{eqnarray}}
\newcommand{\eeqs}{\end{eqnarray}}
\newcommand{\beql}{\begin{equation} \label}

\newcommand{\calV}{{\cal V}}

\newcommand{\parderiv}[2]{\frac{\partial #1}{\partial #2}}
\newcommand{\deriv}[2]{\frac{d #1}{d #2}}

\newcommand{\veps}{\varepsilon}

\newcommand{\p}{\partial}

\newcommand{\scl}{\mathcal{L}}

\newcommand{\imag}{\mathsf{i}}

\usepackage[margin=1in]{geometry}

\newcommand{\ach}[1]{\textcolor{black}{#1}}

\date{}
\begin{document}
\title{Hamilton-Jacobi as model reduction, extension to Newtonian particle mechanics, and a wave mechanical curiosity}

\author{Amit Acharya\thanks{Department of Civil \& Environmental Engineering, and Center for Nonlinear Analysis, Carnegie Mellon University, Pittsburgh, PA 15213, email: acharyaamit@cmu.edu.}}

\maketitle
\begin{abstract}
\noindent The Hamilton-Jacobi equation of classical mechanics is approached as a model reduction of conservative particle mechanics where the velocity degrees-of-freedom are eliminated. This viewpoint allows an extension of the association of the Hamilton-Jacobi equation from conservative systems to general Newtonian particle systems involving non-conservative forces, including dissipative ones. A geometric optics approximation leads to a dissipative Schr\"odinger equation, with the expected limiting form when the associated classical force system involves conservative forces.

\end{abstract}

\section{Introduction}
The Hamilton-Jacobi (H-J) equation of classical mechanics is a nonlinear partial differential equation (PDE) associated with conservative systems equipped with a Lagrangian or an equivalent Hamiltonian \cite{arnold, landau-lifschitz, goldstein1957classical}. It can be viewed as a tool for exactly integrating Hamilton's equations of motion (ordinary differential equations (ODE)), but it is perhaps fair to say that its greater utility has been to provide motivation for formulations of wave mechanics by Schr\"odinger, Bohm, and de Broglie.

In its classical conception, the H-J equation arises from ideas of generating functions in the theory of canonical transformations \cite{arnold, landau-lifschitz, goldstein1957classical} or from the action function (defined from the functional by the same name) \cite[Ch.~9, Sec.~46 C and D]{arnold}, \cite[Ch.~VII, Sec.~43 and 47]{landau-lifschitz}. In this work, we derive the H-J equation as emerging from the question of obtaining a (time-dependent) invariant manifold of the dynamical system representing Newton's equations of motion with conservative forces. This motivation allows a natural generalization of the H-J equation to a system, when non-conservative forces are involved. Using a certain approximation, connection is also made to the Schr\"odinger equation \cite[First Lecture]{schroed_4lect} and a dissipative extension. To our knowledge, this approach to the H-J equation is new and, as will be seen, completely elementary. In particular, it uses no tools from the calculus of variations in the formulation of the problem.

An outline of the paper is as follows: Sec.~\ref{sec:HJ_gen} contains the details of the approach, including the generalization to deal with non-conservative forces in Sec.~\ref{sec:gen_case}, and the association with wave mechanics in Sec.~\ref{sec:wave}. Sec.~\ref{sec:HJ-L} contains a derivation of the H-J equation from one of its usual classical premises, making some connection to the approach in this paper. Sec.~\ref{sec:discuss} contains some concluding remarks.

As for some notational details, we will use $\frac{\p f}{\p (\cdot)_i}, \p_{(\cdot)_i}f$ interchangeably for $(\cdot)_i$ being an argument of any function $f$. When the notation $\p_i f$ is used, the context will be made clear. $t$ will represent time with obvious notation for partial derivatives w.r.t it. The notation $\dot{x}$ for the derivative of the function $t \mapsto x(t)$ will also be used. All velocity and position vector components of particles are assumed written w.r.t a Rectangular Cartesian basis. We will use the summation convention on repeated indices. By an extremal we will mean a trajectory $t \mapsto c(t)$ with derivative represented by $c'$ which satisfies $\frac{\p \scl}{\p c} (c(t), t, c'(t)) - \frac{d}{dt} \left( \parderiv{\scl}{c'}(c(\cdot), \cdot, c'(\cdot))\right)(t) = 0$, where $\scl$ is a given Lagrangian function of the arguments displayed.

\section{\ach{Hamilton-Jacobi equation, an associated wave equation, and H-J-like systems for general Newtonian forces}}\label{sec:HJ_gen}
Consider the Newtonian particle system with prescribed force function \ach{$\hat{f}_i$}, where the index $i = 1, \ldots, 3N$ ranges over the total number of position degrees of freedom with $N$ being the number of particles:
\begin{subequations}\label{eq:newton}
    \begin{align}
        & \dot{v}_i = \ach{\frac{1}{m_i} \hat{f}_i(x,v,t) =: f_i(x,v,t) \qquad (\mbox{no sum on } i)} \label{eq:vel}\\
        & \dot{x}_i = v_i \label{eq:pos}\\
        & v_i(0) = v_i^0; \qquad x_i(0) = x_i^0, \label{eq:ic}
    \end{align}
\end{subequations}
and allow (mass-normalized) forces $f$ for which $\p_{x_j} f_i - \p_{x_i} f_j$ does not necessarily vanish. Also, as shown,  the force functions $f_i$ include the relevant mass of the particle of which $v_i$ is a velocity degree-of-freedom. Suppose now we consider a reduced set of trajectories satisfying the ansatz
\begin{equation}\label{eq:v_ansatz}
v_i(t) = G_i(x(t), t)
\end{equation}
for some functions $G_i$ to be determined.  \ach{For a slow-fast system with $v$ as the fast variables and $x$ slow, and for which the fast dynamics approaches an equilibrium for fixed slow variables (the Tikhonov approach), such an ansatz would be natural without an explicit $t$ dependence in $G$.}

\ach{To derive a set of governing equations for the functions $G_i$, we note that a necessary condition for the ansatz \eqref{eq:v_ansatz} to be valid is that it satisfy \eqref{eq:vel}. This implies that \eqref{eq:G} below must hold. Conversely, let the functions $G_i:\Omega \times (0,T) \to \R$, $\Omega \subset \R^{3n}$ and $(0,T) \subset \R_+$ satisfy the PDE
\begin{equation}\label{eq:G}
    \frac{\p G_i}{\p x_j} (x,t) \, G_j(x,t) + \frac{\p G_i}{\p t} (x,t) - f_i(x,G(x,t), t)  = 0.\footnote{\ach{We note that the sum $\frac{\p G_i}{\p x_j} \, G_j + \frac{\p G_i}{\p t}$ bears a superficial similarity to the material time derivative of the velocity field in continuum mechanics written in the spatial description, were the functions $G$ to be interpreted as such. However, it should be noted that the velocity field for a model of continuum mechanics would be a function on a subset of $\R^3 \times (0,T)$ whereas the $G$ is a function on $\R^{3N} \times (0,T)$, even for a body discretized into a finite number of particles, say $N$. For $N \to \infty$, $G$ is an infinite-dimensional object on an infinite dimensional space; the velocity field of a continuum model, e.g., pressure-less Euler equations, is an infinite dimensional object on a finite dimensional space.}}
\end{equation}
Then, defining `reduced' trajectories by integrating
\begin{equation}\label{eq:red_dyn}
\dot{x}^{(r)}_i (t) = G_i\left(x^{(r)}(t),t\right); \qquad x^{(r)}_{\ach{i}}(0) = x_i^0,
\end{equation}
the accelerations along such trajectories satisfy
\begin{equation}\label{eq:red_traj}
   \dot{v}^{(r)}_i (t) = \ddot{x}^{(r)}_i(t) = f_i\left(x^{(r)}(t), G\left(x^{(r)}(t),t\right), t\right).
\end{equation}
In this sense,} such reduced trajectories satisfy the system \eqref{eq:newton}, but only for velocity initial conditions that satisfy
\begin{equation}\label{eq:red_traj_ic}
v_i^{(r)}(0) = G_i(x^0,0),
\end{equation}
and for $\left(x_i^{(r)}(t), t\right) \in \Omega \times (0,T)$. Of course, several functions, $G^{(I)}$, can be computed for different initial conditions on the velocities (indexed by $I$), making contact with the ideas presented in \cite{sawant2006model} related to computing (overflowing) invariant manifolds of dynamical systems\footnote{We note that due to the dependence on $t$ of the functions $G$, they do not correspond to representations of the standard notion of invariant manifolds, but the notion of `time-dependent' invariant manifolds has also been used in the dynamical systems literature.}. \ach{The approach above for posing \eqref{eq:G} is a straightforward and natural adaptation, to a time-dependent context, of methods for computing invariant manifolds for dynamical systems, see e.g., \cite{sacker1965new,muncaster1983invariant}, with \cite{muncaster1983invariant} abstracting ideas from the Kinetic Theory of Gases \cite{truesdell1980fundamentals} where the functions $G$ are called `Gross Determiners,' with the important distinction from \eqref{eq:v_ansatz} that they do not explicitly depend on time.} We will see that, by invoking a gradient ansatz for the function $G$, as described below, contact is established with the existence of more than one solution to the classical Hamilton-Jacobi equation for Hamiltonian systems, as discussed between \eqref{eq:del_S}-\eqref{eq:leg_trans_1} in Sec.~\ref{sec:HJ-L}.

Suppose now we further restrict the ansatz for the functions $G_i$ by demanding that they be components of the gradient of a scalar valued function $S:\Omega \times (0,T) \to \R$:
\begin{equation}\label{eq:G-S}
G_i(x,t) = \frac{1}{m}\frac{\p S}{\p x_i} (x,t),
\end{equation}
where we assume a common mass $m$ for all particles for notational simplicity. \ach{The potential $S$ has physical dimensions of \emph{Action} $(energy.time)$}. Then \eqref{eq:G} becomes
\begin{equation}\label{eq:plim_hj}
    \begin{aligned}
        & \frac{1}{m^2}\frac{\p S}{\p x_j} \frac{\p^2 S}{\p x_i \p x_j} + \frac{1}{m}\frac{\p^2 S}{\p t \p x_i} - f_i \left(x, \frac{1}{m}\nabla S(x,t), t\right) = 0 \\
        & \Longleftrightarrow \quad \frac{\p}{\p x_i} \left( \frac{1}{2 m^2} |\nabla S|^2 + \frac{1}{m}\frac{\p S}{\p t} \right) (x,t) - f_i\left(x, \frac{1}{m}\nabla S(x,t), t\right) = 0.
    \end{aligned}
\end{equation}
\ach{We note here that the link between \eqref{eq:G} with $f = 0$ and the quadratic Hamilton-Jacobi equation is well-known and much studied in the literature of Optimal Transport, e.g., \cite{villani2021topics}. Of course, our primary interest is in studying systems with general position and velocity dependent force systems.}

Given functions 
\[
F:\R^{3N} \times \R^{3N} \times \R_+ \to \R, \qquad \qquad D_i:\R^{3N} \times \R^{3N} \times \R_+ \to \R
\]
and defining the functions 
\[
V(x) := F(x,0,0), \qquad \qquad  \tilde{F}(x,v,t) := F(x, v,t) - V(x),
\]
now consider force functions of the form
\begin{equation}\label{eq:f_decomp}
  f_i(x,v,t) =  \left( -\frac{1}{m}\frac{\p F}{\p x_i} (x,v,t) +  D_i (x,v,t) \right) =\left(-  \frac{1}{m}\frac{\p V}{\p x_i} (x) -  \frac{1}{m} \frac{\p \tilde{F}}{\p x_i} (x,v,t) + D_i (x,v,t)\right);
\end{equation}
\ach{the chosen `decomposition' of $f$ is motivated by the first term in \eqref{eq:plim_hj} which is a gradient. Of course, a solely position-dependent $V$ is the most common type of force potential that arises in applications related to conservative systems.}

{\color{black} The functions $D_i$ characterize forces that do not arise as gradients, in the space of positions (configuration space), of a scalar potential and include the position-dependent, non-conservative `curl forces' of \cite{berry2015hamiltonian}, as well as dissipative forces. We note that most velocity and time-dependent potentials $\tilde{F}$ result in non-conservative and/or dissipative forces. 

An example of a conservative velocity-dependent potential is as follows: if $v \in \R^{3N}$ is the array of velocity degrees-of-freedom of all $N$ particles and $\mathbb{I}$ is the identity matrix on $\R^{3N}$ and $A \in \R^{3N}$ (used below) is an array required on dimensional grounds, the force (i.e., the negative partial derivative w.r.t $x$) arising from the potential given by $x \cdot \left( \mathbb{I} - \frac{v}{|v|} \otimes \frac{v}{|v|}\right) A$ $\Big($in components $x_i \left(\delta_{ij} - \frac{1}{v_k v_k} v_i v_j\right) A_j \Big)$ is conservative, since it is orthogonal to the velocity at all instants and hence expends no power.}

Using the notation 
\[
\p^1_{x_i} \tilde{F}(x, b(x,t), t) := \lim_{\veps \to 0} \frac{1}{\veps} \left( \tilde{F}(x_1, \dots, x_i + \veps, x_{i+1}, \ldots, x_{3N}, \, b(x,t) \,, t \ ) - \tilde{F}(x, b(x,t), t )\right),
\] 
i.e., a partial derivative w.r.t $x_i$ with `second' argument of $\tilde{F}$ kept fixed, \eqref{eq:plim_hj} then takes the form
\begin{equation}\label{eq:plim_HJ1}
    \frac{1}{m}\frac{\p}{\p x_i} \left( \frac{1}{2m} |\nabla S|^2 + \frac{\p S}{\p t} + V \right) (x,t) + \p^1_{x_i} \tilde{F}\left(x, \frac{1}{m}\nabla S(x,t), t\right) - D_i \left(x, \frac{1}{m}\nabla S(x,t), t\right) = 0,
\end{equation}
and for $F$ of the uncoupled form $F(x,v,t) = F^v(v,t) + F^x(x,0)$ with $F^v(0,t) = 0$, \ach{we have $V(x) := F(x,0,0) = F^x(x,0)$ and so} $\tilde{F}(x,v,t) = F^v(v,t)$ so that $\frac{\p \tilde{F}} {\p x_i} (x,v,t) = 0$, \eqref{eq:plim_HJ1} becomes
\begin{equation}\label{eq:HJ_red1}
 \frac{1}{m} \frac{\p}{\p x_i} \left( \frac{1}{2m} |\nabla S|^2 + \frac{\p S}{\p t} + V \right) (x,t) - D_i \left(x, \frac{1}{m}\nabla S(x,t), t\right) = 0,  
\end{equation}
including the simplest case of $D_i(x,v,t) = - \nu \, v_i, 0 \leq \nu \in \R$.

We note that, formally, any suitably defined continuous solution to (\ref{eq:plim_HJ1}) or (\ref{eq:HJ_red1}) 
defines, through \eqref{eq:G-S} and \eqref{eq:red_dyn}, a reduced, in the sense of (\ref{eq:red_dyn}-\ref{eq:red_traj}-\ref{eq:red_traj_ic}), set of generalized solutions (w.r.t regularity) for $x_i$ in \eqref{eq:newton} corresponding to the specific class of force functions defining (\ref{eq:plim_HJ1}-\ref{eq:HJ_red1}). 

In \eqref{eq:HJ_red1} we recognize the classical Hamilton-Jacobi equation for $D_i = 0$:
\[
  \frac{\p S}{\p t} + \frac{1}{2m} |\nabla S|^2 + V = 0
\]
 (up to an additive arbitrary function of time), \textit{derived from an entirely different premise than classical derivations} based on canonical transformations \cite[Ch.~9]{goldstein1957classical}, or on the action function/Hamilton's principal function \cite[Ch.~9, Sec.~46, C, D]{arnold} and Sec.~\ref{sec:HJ-L}.

Moreover, from this perspective (\ref{eq:plim_HJ1}-\ref{eq:HJ_red1}) may be considered as a consistent generalization of the classical Hamilton-Jacobi equation for conservative  particle dynamics to include a dissipative and non-conservative force. However, with the ansatz \eqref{eq:G-S} in force, it is natural to require $D_i$ in \eqref{eq:HJ_red1} to be a gradient for solutions to exist, and this leaves very few choices for combining $\left(x, \frac{1}{m}\nabla S(x,t), t\right)$ for the desired outcome (noting that $D_i$ cannot be a gradient of a scalar function in its first ($x$) argument by definition \eqref{eq:f_decomp}); we discuss the case of general forces in Sec.~\ref{sec:gen_case}). The simplest such choice is the one related to linear damping mentioned earlier, so that $D = - \frac{\nu}{m} \, \nabla S, \nu \geq 0$, resulting in
\begin{equation}\label{eq:HJ_red2}
\frac{\p}{\p x_i} \left( \frac{1}{2m} |\nabla S|^2 + \frac{\p S}{\p t} + V + \nu S \right) (x,t) = 0 \qquad `\Longleftrightarrow\mbox{'}  \qquad \frac{1}{2m} |\nabla S|^2 + \frac{\p S}{\p t} + V + \nu S = 0, 
\end{equation}
up to an additive function of time which we disregard. \ach{It is interesting to note that the development above provides a physical justification for including a viscosity regularization in the classical H-J equation whose solutions are known to develop shocks in $\nabla S$ from smooth initial conditions.}

\ach{In order to consider particle systems with general, unequal masses, one consider \eqref{eq:vel} written in the form
\[
M_{ij} \dot{{v}}_j = \hat{f}_i(x,v,t),
\]
where $M \in \R^{3N \times 3N}$ is a positive-definite diagonal matrix of masses. Then one can define
\[
\hat{v}_i := M_{ik} v_k
\]
to write \eqref{eq:vel} in the form
\[
\dot{\hat{v}}_i = \hat{f}_i\left(x, M^{-1}\hat{v}, t\right).
\]
One now assumes the ansatze
\[
\hat{v}_i (x,t) = \hat{G}_i(x,t); \qquad \hat{G}_i := \frac{\p \hat{S}}{\p x_i},
\]
which implies
\[
v_i \ = \ M^{-1}_{ik}\hat{G}_k \ = \ M^{-1}_{ik}\frac{\p \hat{S}}{\p x_k}. \qquad (v = M^{-1}\nabla \hat{S}).
\]
We note that the physical dimensions of $\hat{S}$ is also that of $Action$ or $energy . time$.
The corresponding analog of \eqref{eq:G} is then given by
\begin{equation}\label{eq:unequal_mass}
  \frac{\p \hat{G}_i}{\p x_j} (x,t) \, \hat{G}_j(x,t) + \frac{\p \hat{G}_i}{\p t} (x,t) - \hat{f}_i\left(x,M^{-1}\hat{G}(x,t), t\right)  = 0,  
\end{equation}
and that of \eqref{eq:plim_hj} as
\[
\frac{\p}{\p x_i} \left( \frac{1}{2} |\nabla \hat{S}|^2 + \frac{\p \hat{S}}{\p t} \right) (x,t) - \hat{f}_i\left(x, M^{-1}\nabla \hat{S}(x,t), t\right) = 0.
\]
The rest of the development follows in an obvious manner with appropriate adjustment for the single mass $m$ appearing in the earlier simplified development.
}

\subsection{A wave equation corresponding to the scalar Hamilton-Jacobi equation \eqref{eq:HJ_red2}}\label{sec:wave}
\ach{We revert to the equal mass case.} Consider a complex-valued wave function $\Psi: \Omega \times [0,T] \to \mathbb{C}$,
\[
\Psi(x,t) = A(x,t) e^{\imag\frac{S(x,t)}{h}},
\]
where $\imag = \sqrt{-1}$, $A, S: \Omega \times [0,T] \to \R$ are real-valued functions, and $h$ is a constant with physical dimensions of $energy \cdot time$. Then we have the following identities:
\begin{equation}\label{eq:psi_derivs}
    \begin{aligned}
        & S = h \arctan \left( \frac{Im(\Psi)}{Re(\Psi)} \right), \qquad \p_{x_i} S = \frac{h}{2 \imag} \left( \frac{\Psi^* \p_{x_i} \Psi - \Psi \p_{x_i} \Psi^*} {\Psi \Psi^*}\right), \qquad \p_t S = \frac{h}{2 \imag} \left( \frac{\Psi^* \p_t \Psi - \Psi \p_t \Psi^*} {\Psi \Psi^*}\right),
    \end{aligned}
\end{equation}
where we assume that $\arctan: \R \to \left[-\frac{\pi}{2}, \frac{\pi}{2} \right]$ and $Im$ and $Re$ represent imaginary and real parts of their complex-valued argument\footnote{It can be verified that $\p_i S, \p_t S$ as defined are real-valued.}. We also note that $S$ can be equivalently expressed as $S = \frac{h}{2 \imag} \log\left(\Psi/\Psi^*\right)$.

The expressions \eqref{eq:psi_derivs} could be substituted in \eqref{eq:HJ_red2} to yield an equation for $\Psi$ as a `change of variables,' but such an equation would only determine the phase $S/h$ of the wave function and leave the amplitude, $A$, indeterminate. The same observation applies to the equation
\begin{equation}
    \label{eq:psi_HJ}
    \begin{aligned}
        \left(  \frac{1}{2m} |\nabla S|^2 + \frac{\p S}{\p t} + V + \nu S \right) \Psi = 0
    \end{aligned}
\end{equation}
as well, when $S$ and its space-time derivatives are substituted by the corresponding expressions in \eqref{eq:psi_derivs} (we recall that the potential $V$ here is a prescribed function only of $(x)$).

We now follow the interesting observation of Simeonov \cite[Sec.~VIIC]{simeonov2024derivation} that a formal geometric optics \emph{approximation} of the exact nonlinear equation \eqref{eq:psi_HJ} for $\Psi$ produces, for $\nu = 0$, the linear Schr\"odinger equation - an equation from which both the amplitude $A$ and phase $S$ can be determined! Indeed, we have
\begin{equation*}
\allowdisplaybreaks
    \begin{aligned}
        & \p_{x_i} \left( A e^{\imag \frac{S}{h}}\right) = \left( \p_{x_i} A + A \frac{\imag}{h} \p_{x_i} S \right)e^{\imag \frac{S}{h}}\\
        & \p_{x_i x_j} \left( A e^{\imag \frac{S}{h}}\right) = \left( \p_{x_i x_j} A + \p_{x_j} A \, \frac{\imag}{h} \p_{x_i} S + A \frac{\imag}{h} \p_{x_i x_j} S\right)e^{\imag \frac{S}{h}} + \left( \p_{x_i} A + A \frac{\imag}{h} \p_{x_i} S \right) \left(\frac{\imag}{h} \p_{x_j} S\right) e^{\imag \frac{S}{h}}\\
        & \p_{t} \left( A e^{\imag \frac{S}{h}}\right) = \left( \p_{t} A + A \frac{\imag}{h} \p_{t} S \right)e^{\imag \frac{S}{h}},
    \end{aligned} 
\end{equation*}
and in circumstances when, for $\frac{S}{h} \gg 1$, the following approximations are reasonable:
\begin{equation}\label{eq:approx}
\allowdisplaybreaks
    \begin{aligned}
         \frac{h}{\imag} \p_{x_i} \Psi = \frac{h}{\imag} \p_{x_i} \left( A e^{\imag \frac{S}{h}}\right) & \sim (\p_{x_i} S) A e^{\imag \frac{S}{h}} = (\p_{x_i} S) \Psi \\
        \left( \frac{h}{\imag} \right)^2 \p_{x_i x_j} \Psi = \left( \frac{h}{\imag} \right)^2 \p_{x_i x_j} \left( A e^{\imag \frac{S}{h}}\right) & \sim  ( \p_{x_i} S \, \p_{x_j} S) A e^{\imag \frac{S}{h}} = ( \p_{x_i} S \, \p_{x_j} S) \Psi\\
        \frac{h}{\imag} \p_{t} \Psi = \frac{h}{\imag} \p_{t} \left( A e^{\imag \frac{S}{h}}\right) & \sim (\p_{t} S) A e^{\imag \frac{S}{h}} = (\p_{t} S) \Psi,
    \end{aligned} 
\end{equation}
substituting in \eqref{eq:psi_HJ} yields
\begin{equation*}
    \begin{aligned}
        0 & =  \frac{1}{2m} |\nabla S|^2 \Psi + \frac{\p S}{\p t} \Psi + V \Psi + \nu S \Psi \\
        & \sim  \frac{1}{2m}  \left( \frac{h}{\imag} \right)^2 \p_{x_i x_i} \Psi + \frac{h}{\imag} \p_{t} \Psi + V \psi + \nu h \arctan \left( \frac{Im (\Psi)}{Re (\Psi)} \right) \Psi,\\
    \end{aligned}
\end{equation*}
which is rewritten as
\begin{equation}\label{eq:schrod}
    \imag h \frac{\p \Psi}{\p t} \ = \ - \frac{h^2}{2m} \Delta \, \Psi \ + \ V \Psi \ + \ \nu h \arctan \left( \frac{Im (\Psi)}{Re (\Psi)}\right) \Psi,
\end{equation}
where $\Delta = tr(\nabla^2)$ represents the Laplacian operator $\p_{x_i x_i} \Psi = trace(\p_{x_i x_j} \Psi)$, and we note that the last term on the r.h.s of \eqref{eq:schrod} can equivalently be replaced by $\left(\frac{\nu h}{2 \imag} \log\left(\Psi/\Psi^*\right)\right)  \Psi$. Evidently, with the approximation \eqref{eq:approx} in force, the simplest dissipation in the classical mechanics makes the associated wave equation nonlinear. 

It is interesting to note that dissipative extensions of the Schr\"odinger equation have arisen in the literature as described in \cite{gonccalves2018schrodinger} and the discussion of the related literature therein; the development of those ideas proceed along entirely different lines than here.

\ach{In closing this section, we note that the Madelung Transformation (see, e.g.,\cite[Sec.~2.2.1]{bush2020hydrodynamic}, \cite{carles2012madelung}), converts the Schr\"odinger equation to a pair of equations, one of which is the Hamilton-Jacobi equation in the classical limit $h \to 0$ for a particle system without  dissipation \cite{leonard2012schrodinger,khesin2019geometry,von2012optimal}. Here, our main emphasis has been to derive a dissipative Hamilton-Jacobi equation from Newtonian mechanics of particle systems with linear damping. Furthermore, a `converse' of the Madelung route is adopted to arrive at a dissipative Schr\"odinger equation from Newtonian mechanics under specific assumptions, via the latter's dissipative Hamilton-Jacobi equation. More importantly, our primary result is \eqref{eq:HJ_system} in the following Section for which there is no Schr\"odinger equation and a corresponding `Madelung route.'}
\subsection{Hamilton-Jacobi\ach{-like} system for general forces in Newtonian mechanics}\label{sec:gen_case}
As alluded to in the discussion leading up to \eqref{eq:HJ_red2}, not all forces allowed by Newtonian particle mechanics can be associated with a scalar Hamilton-Jacobi equation. In order to deal with general forces, we generalize the ansatz \eqref{eq:G-S} to include a divergence-free component (`orthogonal' to the gradient in an $L^2$ sense):
\begin{equation}\label{eq:G-SR}
G_i(x,t) = \frac{1}{m}\frac{\p S}{\p x_i} (x,t) + R_i(x,t), \qquad \p_i R_i (x,t) = 0.
\end{equation}
Substituting this form in \eqref{eq:G}, using \eqref{eq:f_decomp} and defining $ \tilde{D}_i :=  D_i\left(x, \frac{1}{m} \nabla S + R, t \right) + \frac{\nu}{m} \p_i S$,  we obtain the following \emph{system} of Hamilton-Jacobi\ach{-like}\footnote{\ach{Strictly speaking, \eqref{eq:HJ_system} is a nonlinear second-order quasilinear system. However, since the highest order derivative in $R$ is only first-order and noting the explicit and allowed nonlinearities involving $(\p S, \p R)$ in the second-line of \eqref{eq:HJ_system}, we feel it is justified to refer to \eqref{eq:HJ_system} as a H-J-like system.}} equations for the fields $(S,R)$ corresponding to Newtonian mechanics of mass points under general force systems (using the notation $\p_i = \p_{x_i}$):
{\color{black}
\begin{equation}\label{eq:HJ_system}
\begin{aligned}
    & \frac{1}{m} \p_i \left( \frac{1}{2m} \p_j S \, \p_j S + \p_t S + V + \nu S \right) + \\
    & \quad \frac{1}{m} \big( R_j \p_{ji} S + \p_j R_i \p_j S \big) + \p_i^1 \tilde{F}\left( x, \frac{1}{m} \nabla S + R, t \right) - \tilde{D}_i \left( x, \frac{1}{m} \nabla S + R, t \right)\\
    & \quad + R_j \p_j R_i + \p_t R_i \quad = \quad 0 \\
    & \p_i R_i \quad = \quad 0.
\end{aligned}
\end{equation}
}
 This is a formidable system of equations to solve, and we first note a simplification. One approach for generating a reduced class of `one-way' coupled solutions to \eqref{eq:HJ_system} is to solve \eqref{eq:HJ_red2} and use that solution as forcing in the remaining system for $R$ in \eqref{eq:HJ_system} (i.e., the last three lines of \eqref{eq:HJ_system}). For at most quadratic nonlinearities in the force functions $\p_i^1 \tilde{F}$ and $\tilde{D}_i$ in their second argument, this reduces to a quasilinear system in $R$, up to the divergence-free constraint. 

\section{A derivation of the classical Hamilton-Jacobi equation for conservative particle mechanics}\label{sec:HJ-L}
Having extended the reach of the idea of Hamilton-Jacobi equations from Lagrangian to Newtonian Mechanics, in this Section we give a derivation of the classical Hamilton-Jacobi equation for Lagrangian mechanics. Apart from  the purpose of keeping this work self-contained, the following exercise also shows the necessity of the idea of multiple solutions of the Hamilton-Jacobi equation in order for it to be equivalent in any sense to the classical equations of motion of Lagrangian mechanics for all possible initial conditions admitted by the latter. As a minor remark, the object analogous to Hamilton's principal function or the action function \cite{arnold, landau-lifschitz} is defined from a related, but slightly different, perspective in the following.

Given a Lagrangian $(c,s,c') \mapsto \scl(c,s,c')$, consider the action functional on paths $c:[t_0, t] \to \R^{3N}$ where $t_0$ is a fixed instant of time and $t$ is considered variable:
\[
\hat{S}[c, t] = \int^t_{t_0} \scl\big( c(s), s, c'(s) \big)\, ds; \qquad c'(s) = \frac{dc}{ds} (s).
\]
By invoking a parametrization of physical time as any monotone mapping from a fixed interval $[u_1,u_2] \subset \R$ to the non-negative real numbers, $s: [u_1, u_2] \to \R$ with $\frac{ds}{du} > 0$, the above integral can be equivalently expressed as
\begin{equation}\label{eq:param_hatS}
    \hat{S}[c, s] = \int_{u_1}^{u_2} \scl \left( c(s(u)), s(u), c'(s(u))\right) \, \frac{ds}{du}(u) \, du.
\end{equation}
We would now like to calculate the variation (the directional derivative) of this functional at the state $(c, s)$ in the direction $(\delta c, \delta s)$ (viewed as mappings on the fixed interval $[u_1,u_2]$) where, with some abuse of notation, we will think of all functions as either that of $s$ or $u$, allowed by the monotonicity $s(u)$. The state and the variations will be subject to the following constraints
\begin{equation*}
    \begin{aligned}
        & c(u_1) = x_0; 
        \qquad \delta c(u_1) = 0\\
        & s(u_1) = t_0; 
        \qquad \delta s(u_1) = 0.
    \end{aligned}
\end{equation*}

Then it is shown in  Appendix \ref{app:app1}, \eqref{eq:app1_1}, that
\begin{subequations}\label{eq:del_S}
\allowdisplaybreaks
    \begin{align}
        \delta \hat{S} [c,s; \delta c, \delta s] & = \int_0^t \left( \frac{\p \scl}{\p c} - \frac{d}{ds} \left( \parderiv{\scl}{c'}\right)\right) \cdot \delta c \, ds \notag\\
        & \quad + \int_0^t \left(\parderiv{\scl}{s} - \deriv{}{s} \left( \scl - \parderiv{\scl}{c'} \cdot c'\right)\right) \delta s \, ds \tag{\ref{eq:del_S}}\\
        & \quad +  \left( \scl \Big(c(u), s(u), c'(s(u)\Big) - \parderiv{\scl}{c'}\Big(c(u), s(u), c'(s(u)\Big) \cdot c'(s(u))\right) \delta s(u) \bigg|^{u_2}_{u_1} \notag\\
        &  \quad + \parderiv{\scl}{c'} \Big(c(u), s(u), c'(s(u))\Big) \cdot \delta c(s(u)) \bigg|^{u_2}_{u_1}.\notag
    \end{align}
\end{subequations}
In Appendix \ref{app:app2}, \eqref{eq:app2},  it is shown that for any extremal path in the time interval $(t_0, t)$ - i.e., a path that satisfies the equations of motion of the system and hence for which the integrand of the first integral in \eqref{eq:del_S} vanishes - the second integral in \eqref{eq:del_S} also vanishes. This simply means that along any extremal the rate of change of the Hamiltonian is given by $-\p_s \scl$ and, in particular, if the Lagrangian is independent of an explicit dependence on time, then the Hamiltonian stays constant along any trajectory that satisfies the equations of motion.

Suppose now we group extremal paths starting from $c(t_0) = x_0$ and arriving at $c(t) = x$ on which the functional $\hat{S}$ takes a fixed value, say $S^{(I)}(x,t) \in \R$, for $I$ belonging to some index set; thus, each $I$ corresponds to a different set of extremals departing from $(x_0, t_0)$ and arriving at $(x,t)$, with each extremal in the set resulting in a common value of $\hat{S}$. This protocol then defines a function $S^{(I)}: \R^{3N} \times \R \to \R$, when $x_0$ and $t_0$ are considered fixed for the entire exercise. We will assume that $S^{(I)}$, for each $I$, is a smooth function of $(x,t)$. This amounts to assuming that for fixed $I$ and for each $(x^*, t^*)$ belonging to at least a small neighborhood of $(x,t)$, there exists at least one extremal, say $c^*$, departing from $(x_0,t_0)$ and arriving at $(x^*,t^*)$ for which the function $S^{(I)}$ defined on the neighborhood through $S^{(I)}(x^*,t^*) := \hat{S}[c^*, t^*]$ is smooth\footnote{We note that it is quite possible that there also exists extremals, say $\tilde{c}$, departing from $(x_0, t_0)$ and arriving at $(x^*, t^*)$ for which the value $\hat{S}[\tilde{c}, t^*]$ is significantly different from $S^{(I)}(x,t)$ as well as $S^{(I)}(x^*, t^*)$; such a value for the action may be considered to belong to the range of some other function $S^{(J)}, J \neq I$.}.

The construction also implies that the variation $\delta S^{(I)}(x, t; \delta x , \delta t)$, at $(x,t)$ and in the arbitrary direction $(\delta x, \delta t)$, is given by the variation $\delta \hat{S}[c, s; \delta c, \delta s]$ for $c$ an extremal path for which the action $\hat{S}$ takes the value $S^{(I)}(x,t)$ with $\delta c(s(u_2)=t) = \delta x$, $\delta s(u = u_2) = \delta t$.

Consequently, using \eqref{eq:del_S} we have
\begin{equation}\label{eq:leg_trans_1}
    \parderiv{\scl}{c'} \Big(x, t, c'(t) \Big) = \parderiv{S^{(I)}}{x}(x,t).
\end{equation}
Assuming that the Hessian of the Lagrangian w.r.t its velocity argument, $\p_{c'_i c'_j} \scl (x,t, c')$, is positive-definite for all values of its arguments, any extremal in the set that satisfies $\hat{S}[c,t] = S^{(I)}(x,t)$ while departing from $(x_0,t_0)$ and arriving at $(x,t)$ must have the unique value of $c'(t)$ obtained by inverting the relationship \eqref{eq:leg_trans_1}:
\begin{equation}\label{eq:cprime}
    c'(t) = \left( \parderiv{\scl}{c'} \right)^{-1} \left(x,t, \parderiv{S^{(I)}}{x}(x,t) \right) =: \calV^{(I)}(x,t).
\end{equation}
Here we have introduced the function $\calV^{(I)}$ to represent the fact that the velocities at time $t$ of all such trajectories are a function of $(x,t)$.

Following the argument leading up to \eqref{eq:leg_trans_1} we also have, using (\ref{eq:del_S}-\ref{eq:leg_trans_1}) and for $c'(t)$ given by \eqref{eq:cprime},
\begin{equation}\label{eq:HJ_pre}
\begin{aligned}
      \parderiv{S^{(I)}}{t}(x,t) & =  \scl \Big(x, t, c'(t) \Big) - \parderiv{\scl}{c'}\Big(x, t, c'(t) \Big) \cdot c'(t) \\
      & =  \scl \left(x, t,\left( \parderiv{\scl}{c'} \right)^{-1} \left(x,t, \parderiv{S^{(I)}}{x}(x,t) \right)  \right) - \parderiv{S^{(I)}}{x}(x,t) \cdot \left( \parderiv{\scl}{c'} \right)^{-1} \left(x,t, \parderiv{S^{(I)}}{x}(x,t) \right). 
\end{aligned}
\end{equation}
Defining the Hamiltonian (given a strictly convex Lagrangian, as assumed here):
\[
H(x,t,p) := \min_v \left( p \cdot v - \scl(x,t,v) \right) = p \cdot  \left( \parderiv{\scl}{v}\right)^{-1}(x,t,p)  - \scl\left( x,t, \left( \parderiv{\scl}{v}\right)^{-1}(x,t,p) \right),
\]
\eqref{eq:HJ_pre} then shows that the $S^{(I)}$ function defined above (cf. discussion between (\ref{eq:del_S}-\ref{eq:leg_trans_1})) satisfies the classical Hamilton-Jacobi equation
\begin{equation}\label{eq:HJ}
    \parderiv{S^{(I)}}{t}(x,t) + H \left(x,t,\parderiv{S^{(I)}}{x}(x,t) \right) = 0.
\end{equation}

`Conversely,' it is shown in Appendix \ref{app:app3} that any solution to the Hamilton-Jacobi equation \eqref{eq:HJ} allows the definition of trajectories $t \mapsto c(t)$ which satisfy the classical Lagrangian equations of motion. In particular, from the theory of PDE one may obtain such solutions to \eqref{eq:HJ} by freely specifying initial data $S^{(I)}(x, 0) = g(x)$ or equivalently $\p_x S^{(I)}(x,0) = \p_x g(x) = :v^0(x) \in \R^{3n}$. However, the initial condition on the velocity of the constructed extremals of the action would be subject to the constraint \eqref{eq:cprime1} in which the initial condition on positions $c(0)$ may be arbitrarily specified, with $\calV^{(I)}$ a known function. Thus, a single solution of the H-J equation cannot define all possible classical trajectories emanating from arbitrary `initial' conditions on positions \emph{and} velocities. To obtain the entire class, access to multiple `sheets' $S^{(I)}$, for $I$ varying in its index set (introduced in discussion between \eqref{eq:del_S} and \eqref{eq:leg_trans_1}), is a necessity.

In \cite[Sec.~3.3.2]{evans_PDE} a rigorous derivation of the initial value problem of the H-J equation for an only-momentum dependent Lagrangian is provided, and it is shown that a global solution is given by the minimum of the action characterized by the Hopf-Lax formula \cite[Theorem 6]{evans_PDE}. The perspective here for deriving \eqref{eq:HJ} is slightly different in working with the values of the action functional (for a general Lagrangian) evaluated at critical points, and without any requirement of specifying the value of the action function at any specific instant of time. Of course, as developed herein, the functions $S^{(I)}$ are only defined on local space-time neighborhoods but since \eqref{eq:HJ} is a local statement, this is not an impediment to each such function satisfying the H-J equation.

\section{Concluding remarks and outlook}\label{sec:discuss}
An approach for eliminating the velocity degrees of freedom in the classical mechanics of mass points has been pursued in this work. When the system comprises only conservative forces, the classical Hamilton-Jacobi equation is obtained as the governing equation for a potential function $S^{(I)}$ on the space of particle positions and time. This potential defines a `slaved' velocity function through the relation \eqref{eq:cprime} which, for the standard quadratic Lagrangian in the velocity representing the kinetic energy, reduces to 
\[
\calV^{(I)}(x,t) := \parderiv{S^{(I)}}{x}(x,t),
\]
resulting in a `closed,' i.e., fully defined, reduced dynamics involving only a first-order system for the position degrees of freedom:
\begin{equation}\label{eq:concl_1}
    \deriv{x}{t}(t) = \calV^{(I)}(x(t),t).
\end{equation}
This dynamics allows arbitrary initial conditions only on $x(0)$, with the velocity initial condition slaved to position through \eqref{eq:concl_1} evaluated at $t = 0$. This is in contrast to classical particle mechanics which admits prescribed initial conditions for $x(0)$ and $v(0) = \dot{x}(0)$, this being the price for the reduction. To obtain greater range of prediction with the reduced dynamics, it is in principle conceivable to pre-compute multiple functions/`sheets' $S^{(I)}$ indexed by $I$, as would naturally be the case when solving the Cauchy problem for the H-J equation \eqref{eq:HJ} for a collection of initial data. This approach naturally generalizes to accommodating non-conservative forces in the primal system, resulting in the system \eqref{eq:HJ_system}.

Referring to the Hamilton-Jacobi equation \eqref{eq:HJ} as a many-body problem (the Schr\"odinger equation is a prime example, its independent variables being $(x,t) \in \R^{3n} \times \R$), it is interesting to note that the considerations of this work point to a many-body dynamical formalism for general dissipative dynamical models of continuum mechanics.  Recent work based on duality \cite{CMP18,ach3, V22, acharya2024action} achieve a variational formulation for these, but in a system of dual variables equivalent to the `primal' physical system, resulting in second-order in time evolutions for the dual fields. Because of the variational structure, a Lagrangian (and under suitable circumstances, a Hamiltonian) is available for such problems, even when the `primal' problem is dissipative. The `velocity' degrees-of-freedom of such dual systems can be reduced as demonstrated herein, and a corresponding many-body problem can be defined (assuming a finite dimensional approximation of the problem for simplicity to fit into the ODE based particle model discussed here). Associating the potential $S$ of such a problem with the phase of a wave function, and with the geometric optics approximation, even a Schr\"odinger equation can be associated with such a conservative dual system, corresponding to a non-conservative primal theory of classical fields. Of course, whether such a formalism is of any utility remains to be seen, despite the corresponding success of the association of classical mechanics with quantum mechanics in a similar vein. 

The considerations in Sec.~\ref{sec:HJ-L} require the construction of extremals of the action subject to specified boundary values on only the position at both ends of a specified time interval. This is a non-standard problem statement for solving the equations of motion and the duality technique mentioned above is ideally suited for it \cite{acharya2024action, AG25}.

Also of note is the fact that H-J and quasilinear equations like \eqref{eq:HJ} and \eqref{eq:HJ_system} are some of the hardest nonlinear PDE \emph{systems} to solve, but interesting, nascent progress in the exact realm of defining and obtaining `relaxed' variational dual solutions to such systems is being made, e.g., \cite{CMP18,V22,VA25, sga} - when augmented with (forward and backward) diffusion, of relevance also seem to be the works on Mean Field Game theory \cite{lasry2007mean, Caines2018345}.

Finally, based on the works of \cite{bush2020hydrodynamic, simeonov2024derivation}, there appear to be potential connections of the developments in this work with classical and hydrodynamic quantum analogs proposed therein.
\section*{Acknowledgment}
I thank Janusz Ginster, Ambar Sengupta, Dmitry Vorotnikov, \ach{and Johannes Zimmer} for reading this work and sharing their helpful comments.

\begin{appendix} 
\section{Appendix: Computations in connection to the classical Hamilton-Jacobi equation in Sec.~\ref{sec:HJ-L}}\label{app:HJ_comput}
\subsection{}\label{app:app1}
We compute a variation of the functional shown in \eqref{eq:param_hatS}. In the following, we will omit the inclusion of arguments of integrands in order to avoid excessively cumbersome expressions.
\begin{subequations}
    \allowdisplaybreaks
    \begin{align}
        \delta \hat{S} & = {\int}_{u_1}^{u_2} \left( \left( \p_{c_i} \scl \, \delta c_i + \p_s \scl \, \delta s + \p_{c'_i} \scl \, \delta \left( \deriv{c_i}{u}\deriv{u}{s}\right)\right) \deriv{s}{u} + \scl \deriv{\delta s}{u} \right)\, du \notag\\
        & = \int^{u_2}_{u_1} \left( \left( \p_{c_i} \scl \, \delta c_i + \p_s \scl \, \delta s + \p_{c'_i} \scl \, \deriv{\delta c_i}{u}\deriv{u}{s} + \p_{c'_i} \scl \, \deriv{c_i}{u} \, \delta \left( \deriv{u}{s} \right)\right) \deriv{s}{u} + \scl \deriv{\delta s}{u} \right)\, du \notag
    \end{align}
\end{subequations}
Noting that
\[
\deriv{u}{s}\deriv{s}{u} = 1 \qquad \Longrightarrow \qquad \delta \left( \deriv{u}{s}\right) = - \frac{1}{\left(\deriv{s}{u}\right)^2} \deriv{\delta s}{u},
\]
\begin{subequations}\label{eq:app1_1}
    \allowdisplaybreaks
    \begin{align}
        \delta \hat{S} 
        & = \int^{u_2}_{u_1} \left( \left( \p_{c_i} \scl \, \delta c_i + \p_s \scl \, \delta s  \right) \deriv{s}{u} + \p_{c'_i} \scl \, \deriv{\delta c_i}{u} - \p_{c'_i} \scl \, \deriv{c_i}{s} \, \deriv{\delta s}{u} + \scl \deriv{\delta s}{u} \right)\, du \notag\\
        & = \int^{u_2}_{u_1} \left( \left( \p_{c_i} \scl \, \delta c_i + \p_s \scl \, \delta s  \right) \deriv{s}{u} - \deriv{}{u}\left( \parderiv{\scl}{c'_i}\right) \delta c_i - \p_{c'_i} \scl \, \deriv{c_i}{s} \, \deriv{\delta s}{u} + \scl \deriv{\delta s}{u} \right)\, du \quad + \quad \parderiv{\scl}{c'_i} \delta c_i\bigg|^{u_2}_{u_1} \notag\\
        & = \int^{u_2}_{u_1} \left( \left( \p_{c_i} \scl \, \delta c_i + \p_s \scl \, \delta s  - \deriv{}{s}\left( \parderiv{\scl}{c'_i}\right) \delta c_i\right) \deriv{s}{u}  - \p_{c'_i} \scl \, \deriv{c_i}{s} \, \deriv{\delta s}{u} + \scl \deriv{\delta s}{u} \right)\, du \quad + \quad \parderiv{\scl}{c'_i} \delta c_i\bigg|^{u_2}_{u_1}\notag\\
        & = \int^{t}_{0} \left( \p_{c_i} \scl   - \deriv{}{s}\left( \parderiv{\scl}{c'_i}\right) \right) \, \delta c_i \, ds \quad + \quad \int^{u_2}_{u_1} \left( - \deriv{}{u} \left( \scl - \parderiv{\scl}{c'_i} c'_i\right) + \p_s \scl \, \deriv{s}{u}\right) \, \delta s \, du \notag\\
        & \qquad \qquad + \left( \scl - \parderiv{ \scl}{c'_i}c'_i \right) \delta s \bigg|^{u_2}_{u_1} \quad + \quad \parderiv{\scl}{c'_i} \delta c_i\bigg|^{u_2}_{u_1} \notag\\
        & = \int^{t}_{0} \left( \p_{c_i} \scl   - \deriv{}{s}\left( \parderiv{\scl}{c'_i}\right) \right) \, \delta c_i \, ds \quad + \quad \int^{t}_{0} \left( \p_s \scl - \deriv{}{s} \left( \scl - \parderiv{\scl}{c'_i} c'_i\right)  \right) \, \delta s \, ds \notag\\
        & \qquad \qquad + \left( \scl - \parderiv{ \scl}{c'_i}c'_i \right) \delta s \bigg|^{u_2}_{u_1} \quad + \quad \parderiv{\scl}{c'_i} \, \delta c_i\bigg|^{u_2}_{u_1}. \tag{\ref{eq:app1_1}}
    \end{align}
\end{subequations}
\subsection{}\label{app:app2}
If along any trajectory
\[
\parderiv{\scl}{c_i} - \deriv{}{s} \left( \parderiv{\scl}{c'_i}\right) = 0,
\]
then
\begin{equation}\label{eq:app2}
 \p_s \scl - \deriv{}{s} \left( \scl - \parderiv{\scl}{c'_i} \, c'_i \right) = \p_s \scl - \p_s \scl - \p_{c_i} \scl \, c'_i - \p_{c'_i} \scl \, c''_i + \p_{c'_i} \scl \, c''_i + c'_i \deriv{}{s}\left( \parderiv{\scl}{c'_i} \right) = 0  
\end{equation}
along that trajectory.
\subsection{}\label{app:app3}
Assume a function $S^{(I)}$ has been obtained by solving \eqref{eq:HJ}, so that the function $\calV^{I}$ in \eqref{eq:cprime} is known, and consider a trajectory $t \mapsto c(t)$ generated from integrating
\begin{equation}\label{eq:cprime1}
    \qquad c'_i(t) = \calV_i^{(I)}(c(t),t),
\end{equation}
(cf., (\eqref{eq:red_dyn} and \eqref{eq:G-S}). This means that such a trajectory also satisfies \eqref{eq:leg_trans_1} in the form
\[
\parderiv{\scl}{c'} \Big(c(t), t, c'(t) \Big) = \parderiv{S^{(I)}}{c}(c(t),t) \qquad \mbox{ for all } t.
\]
We will henceforth drop the superscript $(I)$ on $S^{(I)}$ and $\calV^{(I)}$ for notational brevity. Then
\begin{equation}\label{eq:app3_1}
    \deriv{}{t} \left( \parderiv{\scl}{c_i'} \Big(c(\cdot), \cdot, c'(\cdot) \Big) \right) (t) \ = \ \frac{\p^2 S}{\p c_j \p c_i} (c(t), t) \,c'_j (t) \ + \  \frac{\p^2 S}{\p t \p c_i} (c(t), t).
\end{equation}
Using \eqref{eq:HJ_pre}, \eqref{eq:leg_trans_1}, and \eqref{eq:cprime1} we have
\begin{subequations}\label{eq:app3_2}
    \allowdisplaybreaks
    \begin{align}
        \frac{\p^2 S}{\p t \p c_i} (c(t), t) & = \frac{\p^2 S}{\p c_i \p t} (c(t), t) \notag\\
       & =  \left( \parderiv{}{c_i}\left( \scl\big(\cdot, t, \calV(\cdot,t)\big) \ - \ \parderiv{S}{c_j} (\cdot,t) \, \calV_j (\cdot,t) \right) \right)(c(t),t) \notag\\
       & = \parderiv{\scl}{c_i}\big(c(t), t, c'(t)\big) \ + \ \parderiv{\scl}{c'_j}\big(c(t), t, \calV(c(t),t)\big) \parderiv{\calV_j}{c_i}\big(c(t),t\big)\notag\\
       & \quad - \parderiv{^2 S}{c_i \p c_j}\big( c(t), t\big) \, \calV_j (c(t), t) \ - \ \parderiv{S}{c_j}(c(t), t) \, \parderiv{\calV_j}{c_i}(c(t), t) \notag\\
       & = \parderiv{\scl}{c_i}\big(c(t), t, c'(t)\big) \ - \  \parderiv{^2 S}{c_i \p c_j}\big( c(t), t\big) \, c'_j(t),\tag{\ref{eq:app3_2}}
    \end{align}
\end{subequations}
and substituting \eqref{eq:app3_2} in \eqref{eq:app3_1} it is seen that a trajectory defined by \eqref{eq:cprime1} satisfies the equation of motion
\[
\frac{\p \scl}{\p c} \big(c(t), t, c'(t)\big) - \frac{d}{dt} \left( \parderiv{\scl}{c'}\big(c(\cdot), \cdot, c'(\cdot)\big)\right)(t) = 0.
\]
\end{appendix}
\printbibliography
\end{document}